\documentclass[lettersize,journal]{IEEEtran}
\usepackage{amsmath,amsfonts}
\usepackage[T1]{fontenc}
\usepackage{algorithmic}
\usepackage{algorithm}
\usepackage{xcolor}
\usepackage{booktabs}
\usepackage{flushend}

\usepackage{array}
\usepackage{textcomp}
\usepackage{stfloats}
\usepackage{url}
\usepackage{verbatim}
\usepackage{graphicx}
\usepackage{cite}
\makeatletter 
\newcommand{\linebreakand}{%
  \end{@IEEEauthorhalign}
  \hfill\mbox{}\par
  \mbox{}\hfill\begin{@IEEEauthorhalign}
}

\makeatletter
\newcommand*{\rom}[1]{\expandafter\@slowromancap\romannumeral #1@} 
\makeatother 

\begin{document}
\title{A Predictive Flexibility Aggregation Method for Low Voltage Distribution System Control}
\author{Clément Moureau, Thomas Stegen, Mevludin Glavic, Bertrand Cornélusse 
\thanks{The authors are with the University of Liege, Electrical Engineering and Computer Science Department, Liege, Belgium. C. Moureau is a FRIA grantee of the Fonds de la Recherche Scientifique – FNRS}\\
Montefiore Institute, University of Liège, Belgium\\
\{clement.moureau, tstegen, mevludin.glavic, bertrand.cornelusse\}@uliege.be}

\maketitle

\begin{abstract}
This paper presents a method for predictive aggregation of the available flexibility at the residential unit level into a flexibility chart that represents the admissible active and reactive powers, along with the associated flexibility value. The method is also combined with centralized optimization to design a predictive privacy-preserving control scheme to manage low-voltage distribution systems in real-time. Similarly to hierarchical control strategies, this approach divides the optimization horizon into a real-time stage, responsible for decisions in the current market period, and an operational planning stage, which deals with decisions outside of this interval. First, a multiparametric optimization problem is solved offline at the residential unit level. Then, an operational planning problem, also formulated as a parametric optimization problem, is solved to account for the forecasts. The method generates the desired flexibility chart by combining the results of these two problems with measurements. The resulting approach is compatible with real-time control requirements, as heavy computations are performed offline in a decentralized manner. By linking real-time flexibility assessment with energy scheduling, our approach enables efficient and cost-effective management of low-voltage distribution systems. We validate this method on a low-voltage network of 43 buses by comparing it with a fully centralized optimization formulation with perfect foresight and a future-agnostic aggregation method.

\end{abstract}


\section{Introduction}
Although distributed renewable energy sources support the energy transition and help achieve carbon emission reduction targets, their increasing penetration and variability pose significant challenges for the management of distribution networks (DNs). In particular, residential photovoltaic (PV) installations tend to reverse power flow during high-irradiance periods, which may cause network overvoltages. Meanwhile, consumers are accelerating their heating and mobility electrification, increasing network loading and the risk of undervoltage.

Voltage problems will occur more frequently as the penetration of these low-carbon technologies increases. In such situations, expensive measures such as network reinforcement will become necessary \cite{Flex_VS_reinforcement, Frederic_olivier}. Harnessing the flexibility available at the residential level (inverter-interfaced distributed energy resources (DERs), stationary batteries, electric vehicles, and heat pumps) and its efficient real-time control is often proposed as an alternative or complement to reinforcement. 

Even if real-time control of power setpoints may offer an efficient alternative to grid reinforcement, system operators cannot directly actuate these flexibility resources. Due to unbundling requirements, distribution systems operators (DSOs) are generally not allowed to own flexible assets \cite{DSO_flex}, which prevents them from directly actuating and measuring the available flexibility. Instead, they could rely on residential units reporting their real-time flexibility availability and be activated upon request \cite{self_reported_flex, predictive_flexibility1}. Following the work in \cite{self_reported_flex}, we assume that residential units are equipped with energy management systems capable of estimating available flexibility and controlling local assets.

This paper quantifies flexibility using flexibility charts, providing the system operator with an overview of nodal flexibility and the corresponding cost perceived by the residential user. In particular, this work studies how real-time flexibility assessment and energy scheduling can contribute to both defining the cost of flexibility and to efficiently managing DNs. In the following, the term flexibility chart will be used to refer to the aggregated flexibility and its associated cost.

Individual device flexibility charts are usually aggregated into a single diagram to reduce the complexity of coordinating a large number of devices or for data privacy reasons \cite{Aggregation1}. We propose to use multiparametric optimization (MPO) as the first step in aggregating the flexibility \cite{MPC3}. This technique provides a mapping of the optimal solution as a function of varying model parameters by dividing the space of parameters into critical regions. These regions are areas in the parameter space with the same active constraints, and therefore the same functions describing the optimal solutions.

\subsection{Review of the related literature}
Several studies have already explored the concept of harnessing the flexibility of inverter-interfaced generation and loads to improve the management of distribution systems. Most of them focus on the use of PV inverters only \cite{Central1, Central2 }. Other works extended these concepts to loads \cite{Load_shifting}, EVs \cite{EV_flex} or use several of these simultaneously to provide ancillary services to the grid \cite{Residential_LCT}. This opens the door to new opportunities in which prosumers could be compensated for the services they provide to the DN through local electricity markets \cite{Local_markets} or local energy communities \cite{Local_communities}. In \cite{Local_markets}, a local flexibility market is defined as a platform to trade flexibility in local areas. It proposes an optimization problem to manage disconnectable and shiftable loads as well as generators and batteries. 

In \cite{cost_flexibility1}, a method has been proposed to assess the flexibility of buildings by evaluating the amount of energy that can be shifted from one time step to another and the cost resulting from this shift. The method relies on solving multiple optimal control problems, which leads to significant computational complexity. Other methods employ MPC strategies to evaluate the flexibility of buildings under varying electricity tariffs or uncertainties \cite{ predictive_flexibility1, predictive_flexibility2, microgrid_management1, Flex_quantification_EMS}. These approaches quantify the flexibility of residential units to offer cost-effective services to the power system while trying to achieve savings in electricity costs for residential prosumers. In particular, \cite{Flex_quantification_EMS} proposed to account for user preferences in addition to the physical constraints before deriving the available flexibility.

Efficient aggregation and disaggregation methods for active power over a time horizon are proposed in \cite{polymatroid}. A method combining flexibility aggregation through Minkowski sums of zonotopes with the aggregation of piecewise linear cost functions was proposed in \cite{Aggregation1}. Unlike the previous approach, it allows users to assign economic values to their flexibility before aggregation, ensuring cost-effective and economically fair control. Nevertheless, this method does not account for reactive power flexibility and is not suitable for real-time applications.

Real-time aggregation of active and reactive DER flexibility has been investigated in \cite{Flexibility_approx, LCT_voltage_support2} using Minkowski sums and in \cite{Powertech2025} using multiparametric optimization. However, while \cite{Flexibility_approx} and \cite{LCT_voltage_support2} do not account for the cost of flexibility, \cite{Powertech2025} relies on fixed cost functions to penalize photovoltaic curtailment and battery usage, which prevents cost-effective and economically fair control. In \cite{monetarization_flex_chart}, the authors introduce a monetarization step to assess the value of the aggregated flexibility at a given bus. This step consists of evaluating the cost of disaggregation by solving multiple mixed-integer linear programming problems. This brute-force computation of the flexibility charts is not scalable and provides no analytical cost formulation for use at the higher level. 

Despite assessing residential-level flexibility, most of these methods either neglect reactive power flexibility or the cost of activated flexibility.

\subsection{Contributions}

The proposed method relies on MPO to aggregate the residential assets' flexibility and derive the corresponding value function in real-time. Compared to existing works, our approach integrates an operational planning stage formulated as a parametric optimization problem to capture day-ahead forecasts of electricity tariffs, consumption, and production as well as time-coupling constraints. The integration of these forecasts not only enables a more accurate estimate of the cost of activating flexibility but also allows for a fair and optimal flexibility activation. Compared with most strategies proposed in the literature, which penalize deviations from the desired operating point quadratically, the proposed method will first trigger the most cost-effective flexibility and compensate prosumers accordingly.
Furthermore, our formulation is capable of adapting to real-time changes in load, maximum generation, and electricity tariffs, which could pave the way for the provision of real-time ancillary services. Moreover, aggregation enhances privacy, as user preferences are hidden in the aggregated flexibility chart.
The resulting control system remains compatible with real-time requirements, as heavy calculations are performed offline, while being decentralized (at the residential unit level), making it fully parallelizable. A projection step is proposed to derive the real-time flexibility charts from those already calculated offline. In a similar way, an efficient disaggregation process leverages the explicit offline solution obtained through MPO to share setpoints among assets without requiring additional optimization.

To the best of the authors' knowledge, the proposed approach is the first to account for the impact of current decisions on future costs in an aggregation method that includes both active and reactive power flexibility. This paper presents the control architecture in full detail and demonstrates its capabilities using the flexibility of PV systems and BESSs. 

\subsection{Paper organization}
This paper is structured as follows. Section \ref{sect:empc} introduces the general control strategy. Section \ref{sect:offline_MPO} details the formulation of an offline MPO problem that allows aggregation of flexibility and value functions. Section \ref{sect:explicitMPC} presents the MPO formulation of the operational planning problem accounting for forecasts. Section \ref{sect:realtime_control} addresses the real-time projection of the aggregated charts, the central control problem solved by the DSO, and the disaggregation of the resulting setpoints. Numerical simulations demonstrate the effectiveness of the proposed method in Section \ref{sect:results}, while Section \ref{sect:conclusion} concludes.

\section{Control Strategy Overview}
The proposed grid management scheme relies on a centralized controller run by the entity responsible for the distribution network management (i.e., the DSO). This controller coordinates residential units' active and reactive power exchanges with the grid to minimize their operating costs while ensuring safe grid operation. To this end, residential units provide flexibility charts representing both their assets' technical limits and their willingness to deviate from the preferred operating point.

\begin{figure}[h]
    \centering   \includegraphics[width=0.8\linewidth]{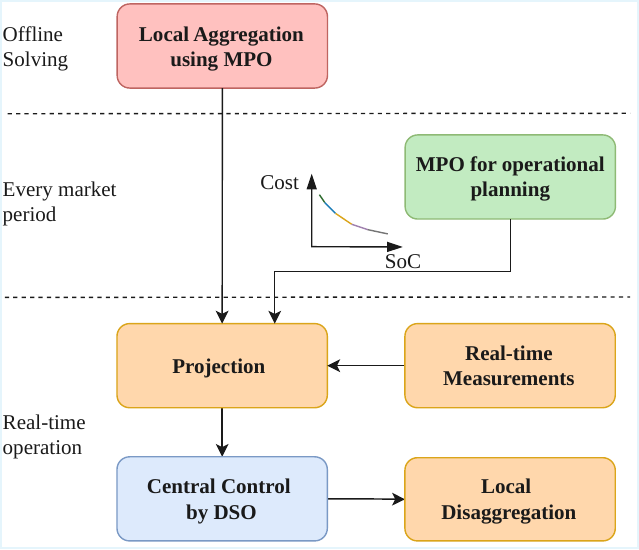}
    \caption{Simplified representation of the different stages of the proposed control strategy}
    \label{fig:MPOOverview}
\end{figure}

Given the fast-changing conditions in distribution systems, setpoints must be updated in real-time, and solving a multi-period problem at each time step to assess the value of flexibility is not tractable. Instead, the determination of a flexibility chart relies on the combination of two types of multi-parametric optimization problems. The key advantage of MPO lies in eliminating repetitive online problem-solving. As parametric programming techniques provide an explicit map of all optimal solutions, changes in operating conditions, either due to a change in operational planning or in ambient conditions, do not require re-optimization. Instead, the optimal solution corresponding to the new value of parameters is already available. This significantly accelerates online evaluation \cite{MPC4}. The three main steps of the proposed approach are summarized in Fig. \ref{fig:MPOOverview}.
\label{sect:empc}

The first MPO is solved offline, once for each residential unit, and provides an explicit solution for real-time dispatch of the residential controllable assets, given: 
\begin{itemize}
    \item an estimation of the current renewable generation potential and consumption of uncontrollable loads
    \item the import and export prices in the current market period
    \item the active and reactive power setpoint desired by the DSO
    \item a function estimating the future cost associated with the system energy state reached at the end of the current market period
\end{itemize}
This problem is solved for a generic market period and provides an explicit mapping between the parameters and the optimal control policies for real-time use. 

The function estimating the cost of the system’s energy state at the end of the current market period is obtained from an operational planning problem formulated as a second MPO. This problem employs day-ahead forecasts of photovoltaic production, load, and electricity prices to compute the cost function, relative to the system’s energy state, by minimizing the total energy cost over the horizon. Building on the approaches in \cite{value_fct_Jonathan} and \cite{value_fct_kumar}, we propose using this function to propagate information from operational planning to the real-time optimization stage. In this formulation, the end of the current market period corresponds to the start of the OP stage, and these two terms are used interchangeably hereafter.

Every $\Delta \tau_t$, the local controllers incorporate the operational planning information and real-time measurements by particularizing the corresponding parameters in the solution of the offline MPO problem. Fixing the values of these parameters allows to express the instantaneous flexibility and corresponding value function in terms of active and reactive power exchanges with the grid. This step can be interpreted as projecting the offline solution onto the $PQ$ plane. 
Using information from all network nodes, the central controller computes the optimal setpoints for each node to minimize the cost of activating flexibility while ensuring safe grid operation. 
Finally, these setpoints are disaggregated among residential assets using the solution of the offline problem.

\section{Local Flexibility Aggregation using MPO}
\label{sect:offline_MPO}

\begin{figure}
    \centering
    \includegraphics[trim=0cm 0.07cm 0cm 0cm, clip, width=\linewidth]{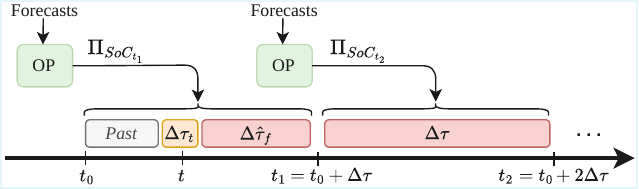}

    \caption{Timeline of the multiparametric optimization problem, separated into three sub-intervals and incorporating a view of the future through $\Pi_{SoC}$}
    \label{fig:MPO_timeline}
\end{figure}

The optimization problem associated with the current market period is formulated as an MPO problem and solved offline for a generic market period and for each residential unit\footnote{In the remainder, lowercase letters are used for variables and uppercase letters for constants. Varying parameters involved in the multiparametric optimization problems are lowercase letters marked with a hat symbol ($\widehat{\phantom{x}}$), uppercase boldface letters denote matrices, and lowercase boldface letters denote column vectors.}. Figure \ref{fig:MPO_timeline} illustrates the timeline adopted by the MPO.

The result of the operational planning problem ($\Pi_{SoC}$) is kept constant within a market period and updated at the end of each period. It maps the future costs to the state of charge (SoC) reached at the end of the period ($t_1$). Each market period is then divided into three intervals. The first represents the past portion of the period. The second corresponds to the real-time response, $\Delta \tau_t$, defined by the central controller update rate (10 seconds). The third interval captures the expected average production and consumption profiles over the following part of the period. The duration of this sub-interval, $\Delta \widehat{\tau}_f$, is treated as a varying parameter and is reduced after each iteration of the central controller to account for time progression. 

The MPO formulation also includes inner polygonal approximations of device flexibility, which depend on varying parameters. These approximations are aggregated by expressing the active and reactive exchange between the node and the grid as parameters, providing a representation of all reachable points of the $PQ$ plane (aggregated capability chart). Aggregation is performed only for the real-time interval. It considers the costs of all previously defined intervals.

\subsection{Objective Function}
The objective is to minimize the daily energy costs that comprise the cost during the current market period ($\Pi_{net}$) and the impact of real-time action on future costs ($\Pi_{SoC})$, reactive power usage ($\Pi_Q$), and battery degradation ($\Pi_{bat}$). 
\begin{equation}
    \min \; \;  \Pi_{net} + \Pi_{SoC} + \Pi_Q + \Pi_{bat}  \label{eq:local_objective}
\end{equation}

The energy cost resulting from power exchanges with the grid during this market period is defined as:
\begin{equation}
    \Pi_{net} = (p_{imp,t}\Delta \tau_t + e_{imp,f}) \widehat{\pi}_{imp} -(p_{exp,t} \Delta \tau_t + e_{exp,f})\widehat{\pi}_{exp}
\end{equation}
with $p_{imp,t}$ and $p_{exp,t}$ representing the imported and exported powers for the next $\Delta \tau_t$ seconds, and $e_{imp,f}$ and $e_{exp,f}$ respectively the imports and exports of energy for the remainder of the market period. $\widehat{\pi}_{imp,t}$ and $\widehat{\pi}_{exp,t}$ represent the import costs and export revenues. These are expressed as varying parameters of the multiparametric problem and updated at each market period.

\begin{figure}[h]
    \centering
    \includegraphics[trim=0cm 0cm 0cm 0cm, clip, width=0.85\linewidth]{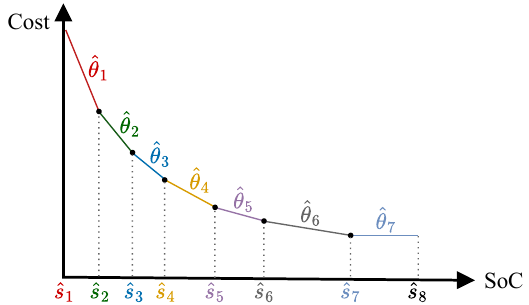}
    \caption{Example of piecewise affine cost function with respect to the SoC at the end of the ongoing market period}
    \label{fig:Cost_approx_new}
\end{figure}

The impact of current actions on future costs is captured through the SoC cost function at the end of the market period, $\Pi_{SoC}$. It is modeled as a piecewise linear function of $N_s$ segments, similar to the one shown in Fig. \ref{fig:Cost_approx_new}. 

\begin{equation}
    \Pi_{SoC} = \sum_{n=1}^{N_s} \widehat{\theta}_{n}\Delta s_n 
\end{equation}
Here, $\widehat{\theta}_{n}$ are the parameters that describe the slopes of each segment. Since the OP is modeled as a linear program, the cost function is convex such that:
\begin{equation}
    \widehat{\theta}_{n}  \leq \widehat{\theta}_{n+1},  \quad \forall n \in \{1,..., N_s-1\}
\end{equation}
The breakpoints defining each segment of the SoC domain are denoted by $\widehat{s}_n$. At the same time, the auxiliary variables $\Delta s_n$ are introduced to determine the cost of the SoC at the end of the current market period, $soc_f$, within the cost function. Equation \eqref{eq:soc_link} establishes the relationship between $soc_f$, the smallest breakpoint $\widehat{s}_0$, and the cumulative increments $\Delta s_n$, thus ensuring consistency.
\begin{align}
    &soc_{f} = \widehat{s}_0+ \sum_{n=1}^{N_s} \Delta s_n \label{eq:soc_link}\\
    &0 \leq \Delta s_n \leq \widehat{s}_{n+1} - \widehat{s}_n, &\forall n \in \{1,..., N_s\} \label{eq:soc_breakpoints} \\
    & \widehat{s}_n \leq  \widehat{s}_{n+1} &\forall n \in \{1,..., N_s\} 
\end{align}
To reduce the number of parameters, the piecewise linear function can be simplified by considering the minimum and maximum SoC values attainable during the market period. Given the SoC at the beginning of the period and the limits on charging and discharging power, many segments of the cost function become unreachable, reducing the number of segments that need to be parametrized.

The choice of $N_s$ reflects a tradeoff between precision and the number of parameters (increasing with $N_s$). If more than $N_s$ regions lie within these reachable SoC bounds, regions with the most similar slopes are merged, at the cost of reduced precision. On the other hand, if fewer than $N_s$ lie within the bounds, additional void regions are introduced to reach $N_s$. In this work, $N_s = 4$. 

Finally, two penalty terms are added. First, the cost associated with instantaneous reactive power usage in the PV ($q_{pv,t}$) and the BESS ($q_{bat,t}$). As shown in \cite{Reactive_Cost} and \cite{ Reactive_Cost2}, reactive power provision reduces the lifetime of inverters and increases losses, such that the resulting cost should be considered. 
\begin{equation}
    \Pi_Q = \pi_Q \left(\lvert q_{bat,t} \rvert + \lvert q_{pv,t} \rvert \right)\Delta \tau_t
\end{equation}
Then, the cost associated with battery degradation due to charging and discharging cycles \cite{Battery_Costs}. 
\begin{equation}
    \Pi_{bat} = \pi_{bat} \left(\lvert p_{bat,t} \rvert \Delta \tau_t + \lvert e_{bat,f} \rvert\right)
\end{equation}

\subsection{Constraints}
\subsubsection{Power Balance}
The power balance constraints define the power exchanges between the grid and the node. For the interval $\Delta \tau_t$, these can be expressed as follows:
\begin{align}
    & p_{exp,t} - p_{imp,t} = p_{pv,t} + p_{dis,t} - p_{ch,t} - \widehat{p}_{load,t}  \label{old_balance} \\
    & p_{exp,t},\; p_{imp,t}, \; p_{ch,t},\; p_{dis,t}, \; p_{pv,t} \geq 0  \nonumber \\
    & q_{exp,t} - q_{imp,t} = q_{pv,t} + q_{bat,t} - \widehat{q}_{load,t} \label{eq:reactivepowerbalance}\\
    & q_{exp,t},\; q_{imp,t} \; \geq 0 \nonumber 
\end{align}
where $ p_{exp,t}$, $p_{imp,t}$, $p_{dis,t}$, and $p_{ch,t}$ denote instantaneous export and import powers and the charging and discharging powers of the BESS, respectively. $\widehat{p}_{load,t}$ is a varying parameter that represents the real-time load and $p_{pv,t}$ denotes the PV power production. Equivalent notations regarding reactive power are used in \eqref{eq:reactivepowerbalance}.

In addition to the previous balances, two parameters, $\widehat{p}_{grid}$ and $\widehat{q}_{grid}$, describe the power exchanges between the node and the grid during the $\Delta \tau_t$ interval. These additional parameters allow to express the instantaneous flexibility and corresponding value functions with respect to active and reactive power exchanges with the grid. They will be specified by the central controller to allow for disaggregation.
\begin{align}
    & \widehat{p}_{grid} = p_{exp,t} - p_{imp,t}  \\
    & \widehat{q}_{grid} = q_{exp,t} - q_{imp,t} 
\end{align}

\subsubsection{Energy Balance}
An energy balance can be performed regarding the following part of the market period.
\begin{align}
    & e_{exp,f} - e_{imp,f} = \widehat{e}_{pv,f} + e_{dis,f} - e_{ch,f} - \widehat{e}_{load,f} \label{new_balance} \\
    & e_{exp,f},\; e_{imp,f}, \; e_{ch,f},\; e_{dis,f} \geq 0   \nonumber 
\end{align}
with most variables defined similarly to those in \eqref{old_balance}. Here, $e$ refers to energy rather than power $p$, and $\widehat{e}_{pv,f}$ denotes the expected PV power production over the remainder of the market period and must therefore be treated as a parameter rather than a variable, with curtailment assumed to occur only upon request from the central controller.

\subsubsection{Power Limits of Assets}
The flexibility limits of inverter-interfaced assets are constrained by the apparent power limits of the inverter, grid codes, and the real-time operating conditions. These feasible regions must be approximated using tailored inner polygonal representations due to the linearity restriction for the constraints of the MPO.

In the case of a PV installation, the operating point $\left(p_{pv,t}, q_{pv,t}\right)$ must satisfy the following constraints, where $\widehat{p}_{\max, pv}$ is a varying parameter that denotes the maximum instantaneous power production of the photovoltaic panels and $S_{nom,pv}$ the inverter's nominal capacity:
\begin{IEEEeqnarray}{rCl}
    p_{pv,t}^2 + q_{pv,t}^2 &\leq& S_{nom, pv}^2, \label{eq:PV_apparent_power} \\
    0 \leq p_{pv,t} &\leq& \widehat{p}_{\max, pv}, \label{eq:PV_MPP} \\
    -p_{pv,t}/3 \leq q_{pv,t} &\leq& p_{pv,t}/3. \label{eq:inverter_lim}
\end{IEEEeqnarray}

Similarly, for a BESS, the constraints are:
\begin{align}
    &p_{bat,t}^2+q_{bat,t}^2 \leq S_{nom, bat}^2 \label{eq:bat_apparent_power}\\
    &soc_{t} = s\widehat{o}c + \left(\eta_{ch} p_{ch,t} -  \dfrac{p_{dis,t}}{\eta_{dis}}\right) \dfrac{\Delta \tau_t}{C_{bat}}  \\
    \text{with, \; }&p_{bat,t} = p_{ch,t} - p_{dis,t} \nonumber\\
    &SoC_{min} \leq  s\widehat{o}c, \; soc_t\leq SoC_{max}  \nonumber
\end{align}
where $\eta_{ch}$ and $\eta_{dis}$ correspond to the charging and discharging efficiencies. The varying parameter $s\widehat{o}c$ represents the current SoC, while $soc_t$ denotes the SoC at the next time step. Both are constrained by the minimum and maximum limits, $SoC_{min}$ and $SoC_{max}$. The battery capacity is $C_{bat}$, with an inverter nominal capacity of $S_{nom,bat}$. 

In the previous equations, (\ref{eq:PV_apparent_power}) and (\ref{eq:bat_apparent_power}) are quadratic and must be linearized to be integrated into the MPO.

\subsubsection{Energy Limits of Assets} 
For the third subinterval, the constraints are expressed in terms of energy rather than power:
\begin{align}
    &0 \leq e_{dis,f}  \leq P_{dis,max} \Delta \widehat{\tau}_f  \\
    &0 \leq e_{ch,f}  \leq P_{ch,max} \Delta \widehat{\tau}_f  \\
     &soc_{f} = soc_{t} + \left(\eta_{ch} e_{ch,f} -  \dfrac{e_{dis,f}}{\eta_{dis}}\right) \dfrac{1}{C_{bat}} \\
     &SoC_{min} \leq  soc_f \leq SoC_{max} \\
    &0 \leq \Delta \widehat{\tau}_f \leq \Delta \tau 
\end{align}
with $\Delta \tau$ being the maximum length of the interval. 

Similarly, the parameter describing the expected power production of the PV panels over the remainder of the market period is constrained by the remaining time and the rated capacity of the PV installation.
\begin{equation}
    0 \leq \widehat{e}_{pv,f}  \leq S_{nom, pv} \Delta \widehat{\tau}_f 
\end{equation}

\subsection{Problem Solving}
The previous problem is solved offline. It provides a set of polyhedral critical regions $\mathcal{R} \in \mathbb{R}^{|\mathcal{P}|}$, where $\mathcal{P}$ is the set of varying parameters. 
Each polyhedron ($r$) is defined by a set of linear inequalities (\ref{eq:Axb}) and is characterized by its associated value function $f_r: = \mathbb{R}^{|\mathcal{P}|} \mapsto \mathbb{R}$ (\ref{eq:costs}) and linear optimal policy $\mathbf{G_r}: = \mathbb{R}^{|\mathcal{P}|} \mapsto \mathbb{R}^{|\mathbf{u_r^*}|}$ (\ref{eq:policy}). These regions, along with their corresponding policies and value functions, are stored locally to be used in real-time. 
\begin{align}
&\mathbf{A_{r}}\mathbf{x} \leq \mathbf{b_{r}},  \quad \; \; \;\forall r \in \mathcal{R} \label{eq:Axb} \\
& \Pi^*_r = f_r(\mathbf{x}), \quad \;\forall r \in \mathcal{R} \label{eq:costs} \\
&\mathbf{u_r^*} = \mathbf{G_r(\mathbf{x})}, \quad  \forall r \in \mathcal{R} \label{eq:policy} 
\end{align}
where $\mathbf{x}$ is the vector that collects the varying parameters and $\mathbf{u_r^*}$ the vector of optimal decision variables.

\section{Operational Planning Formulation} 
\label{sect:explicitMPC}
\label{sect:Operationnal_planning}

This problem yields the cost-to-go function that enters into the problem formulated for the real-time operation.
Hence, every market period, a parametric optimization problem is solved using updated forecasts. This computation provides a mapping of the SoC cost at the beginning of the OP stage, showing how variations in SoC influence system costs under the forecasted load and production profiles. To capture daily production and consumption patterns, a one-day horizon is divided into $N_C$ intervals.

\subsection{Objective and Constraints}
The objective minimizes both energy costs and battery degradation. Instead of computing the cost over the interval as the sum of import costs and export revenues, we adopt a more realistic and conservative approach. Let $E_{load,net,t}$ and $E_{pv,net,t}$ denote the expected net load and net PV production over the interval, respectively. Here, net load (resp. PV production) denotes the portion of the load (resp. PV production) that is not simultaneously covered by PV production (resp. load) during the interval. This allows us to express the expected grid costs as the sum of $c_{ch,t}$ and $c_{dis,t}$, which represent the respective impacts of charging and discharging on grid costs over the considered time interval.
\begin{subequations}
\begin{align}
   \min &\sum_{t=1}^{N_C}  c_{ch,t} +  c_{{dis},t} +  \pi_{bat} (e_{{ch},t} + e_{dis,t})
    \label{eq:MPC_obj}\\
\text{s.t.} & \;  \forall t  \in \{1, \dots, N_C\}, \nonumber\\
    &\Pi_{exp,t}  (E_{{load,net},t} - e_{{dis},t})  \leq c_{{dis},t}  \\    &\Pi_{imp,t}  (E_{{load,net},t} - e_{{dis},t})  \leq c_{{dis},t}  \\
    &\Pi_{exp,t}  (e_{{ch},t} - E_{{pv,net},t})  \leq c_{{ch},t}  \\      &\Pi_{imp,t}  (e_{{ch},t} - E_{{pv,net},t})  \leq c_{{ch},t}  \\ 
    &0 \leq e_{{dis},t}, e_{{ch},t} \\
    &e_{{dis},t} + e_{{ch},t} \leq S_{nom, bat} \Delta \tau_t\\
    &soc_{t} = soc_{t-1} + \left(\eta_{ch} e_{ch,t} -  \dfrac{e_{dis,t}}{\eta_{dis}}\right) \dfrac{1}{C_{bat}} \\
    & SoC_{min} \leq  soc_{t-1}  -  \dfrac{e_{dis,t}}{\eta_{dis} C_{bat}} \label{eq:SOC_min_cons}\\    
    & soc_{t-1}  +  \dfrac{\eta_{ch} e_{ch,t}}{ C_{bat}} \leq SoC_{max} \label{eq:SOC_max_cons}\\
    & soc_{0} = s\widehat{o}c_0 
\end{align}
\end{subequations}
where $e_{ch,t}$ and $e_{dis,t}$ denote the amount of energy charged and discharged in the battery during the intervals. $\pi_{bat}$ represents the cost associated with battery degradation. The battery SoC at the end of each interval is $soc_t$, while the varying parameter $s\widehat{o}c_0$ represents the SoC at the beginning of the OP stage. $\Pi_{imp,t}$ and $\Pi_{exp,t}$ are the import and export costs, and $\Delta \tau_t$ indicates the interval length.

This formulation better captures the actual dynamics of the BESS SoC, in particular by accounting for its charging and discharging cycles over the future horizon, which improves the accuracy of the SoC cost function. Moreover, constraints \eqref{eq:SOC_min_cons} and \eqref{eq:SOC_max_cons} are intentionally conservative. They ensure that, regardless of when charging or discharging occurs within the interval, the BESS's SoC remains within its operational limits.

Solving this problem yields a set of critical regions in $\mathbb{R}$, $s\widehat{o}c_0$ being the only varying parameter. Each region is defined by inequalities that specify the values of $s\widehat{o}c_0$ for which it is active and is characterized by a piecewise affine cost function (an example is shown in Fig. \ref{fig:Cost_approx_new}).

\section{Real-time Control}
\label{sect:realtime_control}
In real-time, all local controllers communicate their instantaneous value functions to the central controller. As detailed in the previous sections, the solution of the offline MPO problem provides the optimal system costs $\Pi^*$ and decision variables $\mathbf{u^*}$ as functions of the varying parameters. Most of these can be fixed by:
\begin{itemize}
    \item measuring the system parameters ($s\widehat{o}c$, $\widehat{p}_{load}$, $\widehat{q}_{load}$ and $\widehat{p}_{max,pv}$),
    \item forecasting the PV production and load for the remaining of the 15-minute interval ($\widehat{e}_{pv}$, $\widehat{e}_{load}$),
    \item imposing the cost function computed by the OP stage ($\widehat{s}_n$, $\widehat{\theta}_n$),
    \item and considering the electricity tariffs and the remaining time in the market period ($\widehat{\pi}_{exp}$, $\widehat{\pi}_{imp}$, and $\widehat{\tau}_f$).
\end{itemize} 

As a result, the value function can be expressed solely in terms of active and reactive power exchanges with the grid. After collecting the flexibility charts from all nodes, the central controller computes the optimal nodal setpoints and communicates them to the local controllers responsible for disaggregation.

\subsection{Online Flexibility Assessment Through Projection}
\label{sect:flex_and_cost}

The offline MPO divides the parameter space into regions, each characterized by different value functions. By definition, a region is active if it contains the point defined by all varying parameters. Although this point remains unknown at this stage, the set of potentially active CRs can be restricted to those that are consistent with the fixed parameter values. To identify this set, it is necessary to locate all regions in $\mathcal{R}$ that intersect the subspace described after fixing the values of the parameters: $\widehat{s}_n$, $\widehat{\theta}_n$, $\widehat{p}_{load}$, $\widehat{q}_{load}$, $s\widehat{o}c$, $\widehat{p}_{max,pv}$, $\widehat{\pi}_{exp}$, $\widehat{\pi}_{imp}$ and $\Delta \widehat{ \tau_r}$.

Each CR is defined by a constraint set $\mathbf{A_{r}}\mathbf{x} \leq \mathbf{b_{r}}$. Fixing some parameters in $\mathbf{x}$ allows subtracting their contributions to $\mathbf{b_r}$ using the associated columns in $\mathbf{A_r}$. 

Without loss of generality, we assume that the first $i$ elements of $\mathbf{x}$ correspond to fixed parameters. The associated subvector is denoted $\mathbf{\Tilde{x}}$, and the submatrix formed by the corresponding columns of $\mathbf{A_r}$ is denoted by $\mathbf{\Tilde{A}_{r}}$. The remaining elements of $\mathbf{x}$ are collected in $\bar{\mathbf{x}}$ while the remaining columns of $\mathbf{A_r}$ are stored in $\mathbf{\bar{A}_r}$. This allows the reduced system to be written as:
\begin{equation}
    \mathbf{\bar{A}_r} \mathbf{\bar{x}} \leq \mathbf{\bar{b}_r}
\end{equation}
where $\mathbf{\bar{b}_r} = \mathbf{b_r} - \mathbf{\Tilde{A}_{r}}\mathbf{\Tilde{x}}$

This results in a reduced linear system with fewer parameters. However, the number of critical regions remains the same. To reduce it, rows of the resulting matrix $\mathbf{\bar{A}_r}$ with zero coefficients are separated from the others. The system feasibility is then verified by checking if the corresponding entries in $\mathbf{\bar{b}_r}$ are positive. If any of them are negative, at least one inequality in $\mathbf{\bar{A}_r} \mathbf{\bar{x}} \leq \mathbf{\bar{b}_r}$ cannot be satisfied for any values of $\mathbf{\bar{x}}$ (of $\widehat{p}_{grid}$ and $\widehat{q}_{grid}$). Otherwise, the new constraint set defines a polyhedron in $\mathbb{R}^2$ that is potentially an active region. 
\begin{equation}
    \mathbf{\bar{A}_{r}} \begin{bmatrix} \widehat{p}_{grid} \\ \widehat{q}_{grid}  \end{bmatrix} \leq \mathbf{\bar{b}_r} \label{eq:free_part}
\end{equation}

Iterating over all regions in $\mathcal{R}$, a set of polygonal critical regions is obtained. We denote this new set as $\mathcal{\bar{R}}$.

\begin{figure}[h]
    \centering
    \includegraphics[trim=3.6cm 2.5cm 4.7cm 3.5cm, clip, width=0.85\linewidth]{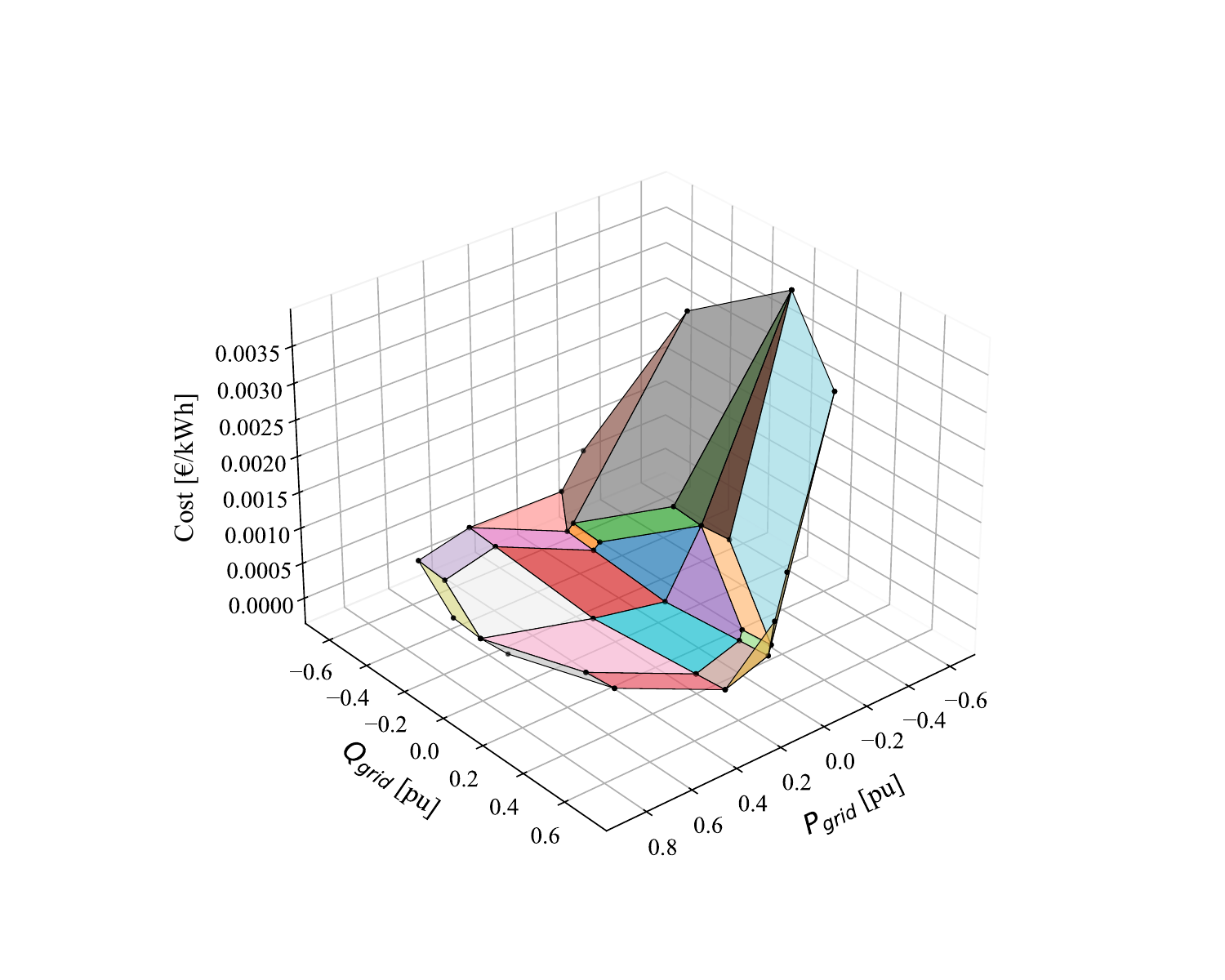}
    \caption{Value function in the $PQ$ plane for example of parameters}
    \label{fig:triangulation}
\end{figure}

Figure \ref{fig:triangulation} represents an example of such a set with $\widehat{p}_{grid}$ and $\widehat{q}_{grid}$ being the only two remaining parameters. Each region is associated with a linear value function, such that an aggregate value function $f(p_{grid},q_{grid})$ is defined. These critical regions and corresponding value functions are communicated to the central controller.

\subsection{Central Optimization}
\label{sect:central_controller}
The central controller is responsible for computing the optimal power setpoints for each node based on their aggregate value functions. The controller relies on the \textit{Branch Flow Model} \cite{BranchFlow} to represent a radial low voltage DN. In the following formulation, let $\mathcal{N}$ represent the set of network nodes and $\mathcal{E}$ the set of directed lines.

\subsubsection{Objective}
The objective function is formulated as the minimization of the sum of the local and loss costs:
\begin{equation}
    \min _{p, q} \sum_j f_j\left(p_j, q_j\right) + \pi_l \sum_{(ij)\in \mathcal{E}} R_{ij}l_{ij}
    \label{eq:total_cost}
\end{equation}
where $p_j$ and $q_j$ are the active and reactive power injections at node $j$ and $f_j$ the corresponding value function, similar to the one shown in Fig. \ref{fig:triangulation}. $\pi_l$ is the cost associated with network losses, $R_{ij}$ is the resistance of the line from node $i$ to node $j$ and $l_{ij}$ the squared current magnitude in that line. Since $f_j$ is piecewise linear and convex, it can be represented using a set of linear constraints, avoiding the introduction of binary variables.\\

\subsubsection{Branch Flow Constraints} These describe the coupling between nodal voltages and power exchanges.
\begin{align}
& p_j=\sum_{(jk) \in \mathcal{E}} p_{jk}-\sum_{(ij) \in \mathcal{E}}\left(p_{ij}-R_{ij} l_{ij}\right), \quad \forall j \in \mathcal{N}, \label{eq:centralized_active_power}\\
& q_j=\sum_{(jk) \in \mathcal{E}} q_{jk}-\sum_{(ij) \in \mathcal{E}}\left(q_{ij}-X_{ij} l_{ij}\right), \quad \forall j \in \mathcal{N}, \label{eq:centralized_reactive_power}\\
& v_j=v_i-2\left(R_{ij} p_{ij}+X_{ij} q_{ij}\right)+\left(R_{ij}^2+X_{ij}^2\right) l_{ij}, \; \forall(i, j) \in \mathcal{E}, \label{eq:centralized_voltage1}\\
& l_{ij} v_i = p_{ij}^2+q_{ij}^2, \quad \forall(i, j) \in \mathcal{E}
\label{eq:centralized_line_current}
\end{align}
where $v_j$ is the squared voltage magnitude at node $j$, $p_{ij}$ and $q_{ij}$ denote the active and reactive power flows in the line from node $i$ to node $j$, and $X_{ij}$ is the corresponding reactance. \\

\subsubsection{Asset Feasible Set}
The aggregated flexibility of all local assets defines the feasibility set. 
\begin{equation}
     (p_j , q_j) \in \mathcal{U}_j,  \quad \forall j \in \mathcal{N}
\end{equation}

where $\mathcal{U}_j$ is the union of the polygons in $\mathcal{\bar{R}}_j$. \\ 

\subsubsection{Network Constraints}
Voltage levels must stay within a predefined range. This constraint is defined as:
\begin{equation}
    0.95^2 \leq v_{j} \leq 1.05^2, \quad \forall j \in \mathcal{N} 
\end{equation}

\subsection{Disaggregation of Power Setpoints}
\label{sect:disaggregation}

The disaggregation process follows a structure similar to that of the flexibility assessment. For each residential unit, the central controller computes and communicates power setpoints ($p_j, q_j$) that represent the optimal power exchange at a given node. These setpoints must be cost-effectively shared among the local assets. This is performed by fixing the remaining varying parameters of $\mathbf{\bar{x}}$ in (\ref{eq:free_part}), $\widehat{p}_{grid}$ and $\widehat{q}_{grid}$. This enables the identification of the region in which the desired operating point lies. Mathematically, this comes back to identifying the region in $\mathcal{\bar{R}}$ for which constraint \eqref{eq:free_part} is satisfied.

The local cost is evaluated using (\ref{eq:costs}) within the corresponding critical region, and the optimal policy from (\ref{eq:policy}) specifies the setpoints of each asset. This step requires only a function evaluation rather than solving an optimization problem.

\subsubsection*{Remark}This work implicitly assumes that the DSO compensates residential users, providing flexibility to the grid if they are asked to deviate from their preferred operating point. This opens the door to gaming behaviors from residential users, who might shape their charts to maximize compensations. Some rules should thus potentially be set to avoid this behavior, but this is out of the scope of our present work. 

\section{Case Study and Numerical Results}
\label{sect:results}
\begin{figure}[ht]
     \centering
         \centering
         \includegraphics[trim=0cm 0.8cm 0cm 0.6cm, clip, width=0.99\linewidth]{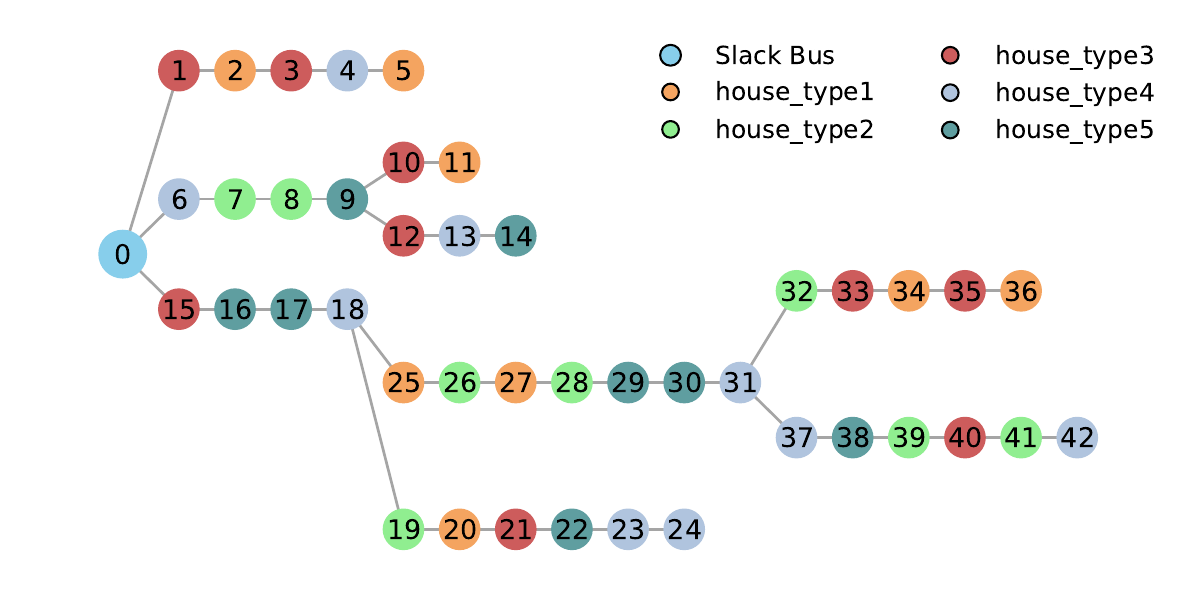}
         \label{fig:Studied_network_new}
         \caption{Studied 43-node suburban system from the SimBench library}
\end{figure}

We consider the low-voltage 43-node DN represented in Fig. \ref{fig:Studied_network_new}. This network model, derived from \textit{SimBench} \cite{simbench}, has been modified by assigning each network node a randomly selected residential unit from Table \ref{tab:building_types} and a random load profile.

\begin{table}[H]
    \centering
    \caption{Typical residential units}
    \label{tab:building_types}
    \begin{tabular}{r r r r}
        \toprule
        \textbf{House Type} & $P_{max,pv}$ [kWp] & $C_{bat}$ [kWh] & $S_{nom,bat}$ [kVA] \\ \midrule
        1 & 0 & 0 & 0 \\ 
        2 & 10 & 0 & 0 \\
        3 & 5 & 0 & 0  \\
        4 & 10 & 30 & 10 \\ 
        5 & 5& 20 & 5 \\ \bottomrule
    \end{tabular}
\end{table}

The load and the maximum available PV production are updated every 10 seconds, which corresponds to the central controller's solving rate. Load profiles are derived from a dataset providing measurements of household active and reactive power consumption \cite{load_data}. Given the proximity of the nodes, the same irradiance curve is assumed for all PV systems. This profile is obtained from real PV production measurements in Belgium on July 27, 2025. Import and export tariffs are based on the electricity spot prices observed in Belgium on the same day. These tariffs are then adjusted to account for transport and distribution costs and taxes. 

The parametric problems are formulated using the MPT3 library from Yalmip \cite{Yalmip} in \textsc{MATLAB}. The central optimization problem is formulated using Pyomo \cite{pyomo} and solved with IPOPT \cite{ipopt}.

The performance of the proposed predictive flexibility aggregation method (PFA) is assessed by comparing it with an optimal scenario in which an omniscient controller (OMNI) simultaneously optimizes all setpoints in a day-ahead planning framework. Such a strategy serves as an ideal reference scenario but is impractical for real-world implementation due to uncertainty in day-ahead forecasts and long solving time. Additional comparisons are performed with the control method presented in \cite{Powertech2025}, which employs a similar aggregation method but relies solely on the current system state (FA).

\begin{figure*}
    \centering
    \includegraphics[trim=0cm 0.5cm 0cm 0.5cm, clip, width=\linewidth]{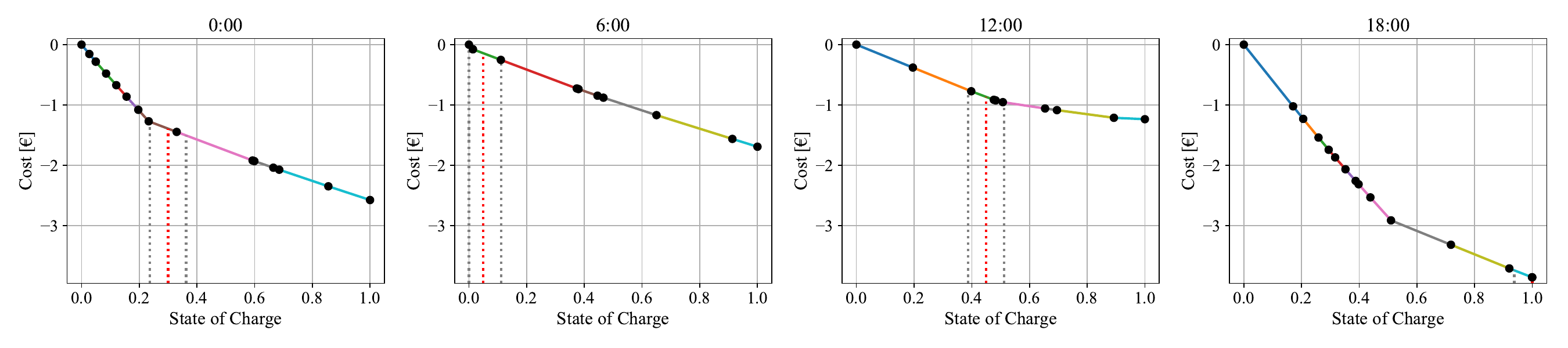}
    \caption{Output of the OP stage for node 9 at 4 different time periods and SoC values of 0.3, 0.05, 0.45, and 1, respectively.}
    \label{fig:OP_results}
\end{figure*}

\begin{figure*}
    \centering
    \includegraphics[trim=0cm 0.05cm 0cm 0.5cm, clip, width=\linewidth]{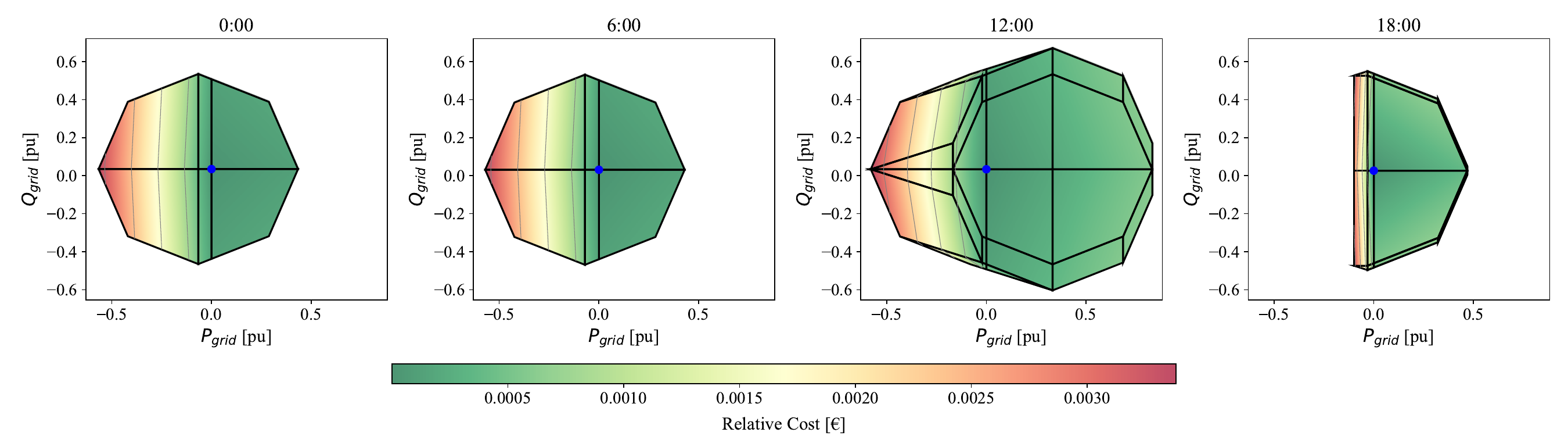}
    \caption{Output of the real-time projection step for node 9 at 4 different times of the day with the contour lines in grey and the different critical regions separated by black lines.}
    \label{fig:PQ_charts_cost}
\end{figure*}
\subsection{Intermediate Results Interpretation} 
To better understand the different stages of the proposed approach, the intermediate results obtained at each stage are detailed for four periods of the day. 
\subsubsection{Operational Planning Stage}
Figure \ref{fig:OP_results} presents the results of the OP stage, in which all costs are expressed relative to an empty BESS. As expected, the value assigned to the SoC is higher when the following intervals are characterized by high load and low generation capacity. Thus, at 6 p.m. and midnight, the initial segments exhibit a steep slope. Then, once the SoC is sufficient to avoid any future imports, a significant decrease in the slope is observed. Similarly, at noon, the expected renewable generation is sufficient to prevent any future imports, resulting in a low SoC value.

Figure \ref{fig:OP_results} also shows the current SoC of the BESS (in red) as well as the maximum and minimum values that could be reached at the end of the current market period (in grey). This illustrates how many segments should be considered in $\Pi_{SoC}$ to guarantee no precision loss. It also shows that approximating $\Pi_{SoC}$ with only four segments ($N_s = 4$) results in minimal precision loss when merging segments, as many of them have similar slopes.
\subsubsection{Real-time Stage}
Figure \ref{fig:PQ_charts_cost} presents the flexibility charts obtained after the projection step. To facilitate analysis, costs are expressed relative to the point where active power exchange and reactive power compensation are zero. This point is marked in blue on each chart. 
At midnight and 6 a.m., the irradiance is zero, and the flexibility chart corresponds to that of the battery alone. These two graphs are also very similar in terms of relative costs, as the electricity tariffs are nearly identical, and as the respective SoCs are located in regions with similar slopes in Fig. \ref{fig:OP_results}. At noon, high irradiance results in significantly greater available flexibility and a larger number of regions. In contrast, the smallest flexibility chart is observed at 6 p.m, when the BESS has reached its maximum SoC and irradiance is low. At this time, the battery can only discharge, and most of the flexibility lies on the export side of the graph.
It is also worth noting that contour lines are piecewise linear, given that the cost function is itself piecewise linear.

\subsection{Simulation Results} 
Optimizing all setpoints within a day-ahead planning framework with a 10-second simulation time step proved computationally intractable. The omniscient approach requires prohibitively long computation times due to the large number of decision variables, which are coupled by both network and SoC constraints. 
To estimate the optimal performance bound, the OMNI was formulated with a simulation time step three times larger, and still required 62 hours to solve. Although the results in this setting differ from those of the original problem, they provide an estimate of the achievable performance.

\begin{table}[h]
    \centering
    \caption{Power output comparison for different control strategies}
    \begin{tabular}{lrrr}
        \toprule
        & \textbf{OMNI} & \textbf{FA} & \textbf{PFA} \\ \midrule
        Cost (€) & 38.79 & 74.11 & 42.88 \\ 
        Corrected Cost (€) & 38.79 & 57.69 & 41.93 \\ 
        $E_{prod,pv}$ (kWh) & 1486.37 & 1541.07 & 1480.89 \\ 
        $E_{ch,bat}$ (kWh) & 463.03 & 396.23 & 443.49 \\ 
        $E_{dis,bat}$ (kWh) & 438.28 & 166.51 & 411.78 \\ 
        $Q_{prod,tot}$ (kvarh) & 479.18  & 694.10 &  520.58 \\ \bottomrule
    \end{tabular}
    \label{tab:Results_comp43}
\end{table}

Table \ref{tab:Results_comp43} summarizes the active and reactive production of the assets and the total daily energy costs. It demonstrates that the proposed method improves the results obtained with FA and that, as expected, the OMNI exhibits the lower costs. In Table \ref{tab:Results_comp43}, two costs are proposed. The first represents the actual cumulative cost for prosumers, while the second adjusts the proposed cost to account for the differences in final SoCs by valuing the surplus of stored energy at the next morning's day-ahead market price.

\begin{figure}[h]
    \centering
\includegraphics[trim=0cm 0.6cm 0cm 0.6cm, clip, width=0.99\linewidth]{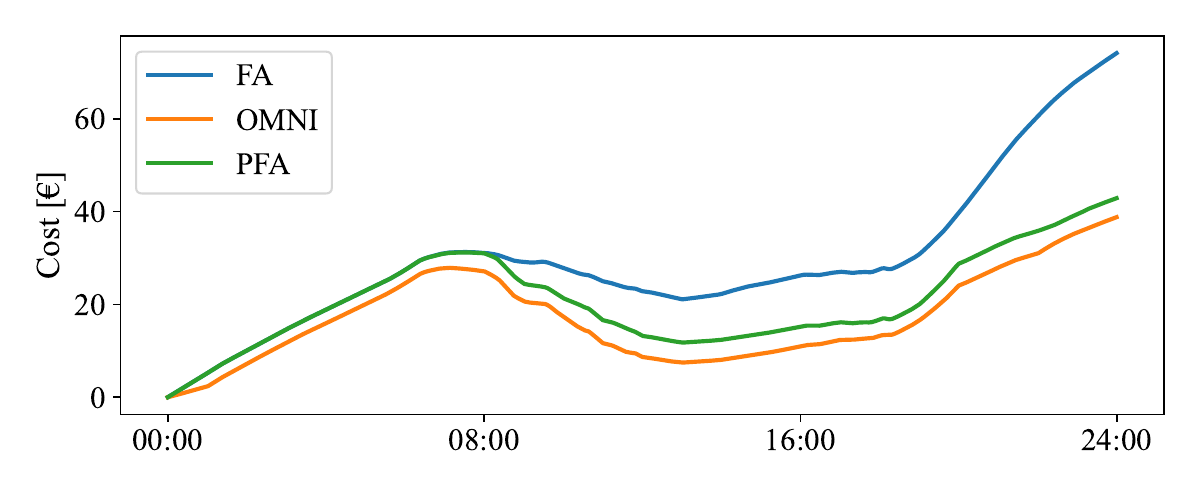}
    \caption{Evolution of the total system cost presented in Eq. \ref{eq:total_cost} over the day}
    \label{fig:Cost_cumsum}
\end{figure}

The evolution of the total cumulative costs is shown in Fig. \ref{fig:Cost_cumsum}. After an initial drift, mainly due to the anticipative discharge of the BESSs in the OMNI, both OMNI and PFA strategies appear to evolve along parallel trajectories, suggesting that the proposed method tries to track the optimal trajectory.

\begin{figure}[h]
    \centering
    \includegraphics[trim=0cm 0.7cm 0cm 0.7cm, clip, width=0.99\linewidth]{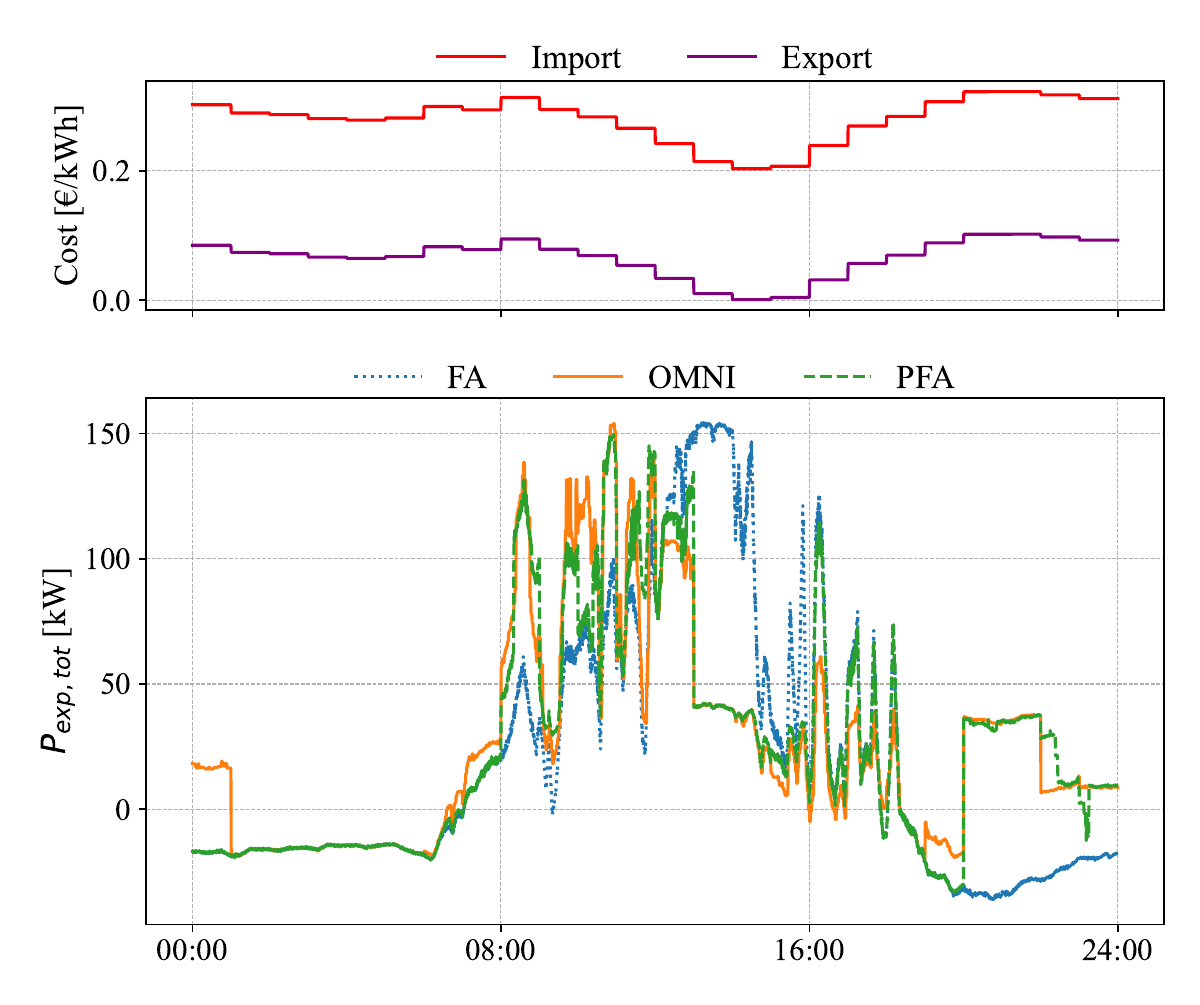}
    \caption{Import and export tariffs and total exported power comparison for the three proposed control methods}
    \label{fig:E_exp}
\end{figure}

Regarding PV production, FA achieves the least curtailment. This strategy uses fixed import and export tariffs for the whole day, which, during periods of low electricity prices, are significantly higher than the actual tariffs. This overestimation tends to favor the excessive use of reactive power to maximize photovoltaic production. As expected, the OMNI controller achieves less curtailment than the PFA since it optimizes all operating points simultaneously and can therefore anticipate potential constraint violations. This anticipation is very well reflected in Fig. \ref{fig:E_exp}, where the omniscient strategy begins exporting immediately, knowing that there might be overvoltages in future periods of high export tariffs.

\begin{figure}[h]
    \centering
    \includegraphics[trim=0cm 0.7cm 0cm 0.5cm, clip, width=0.99\linewidth]{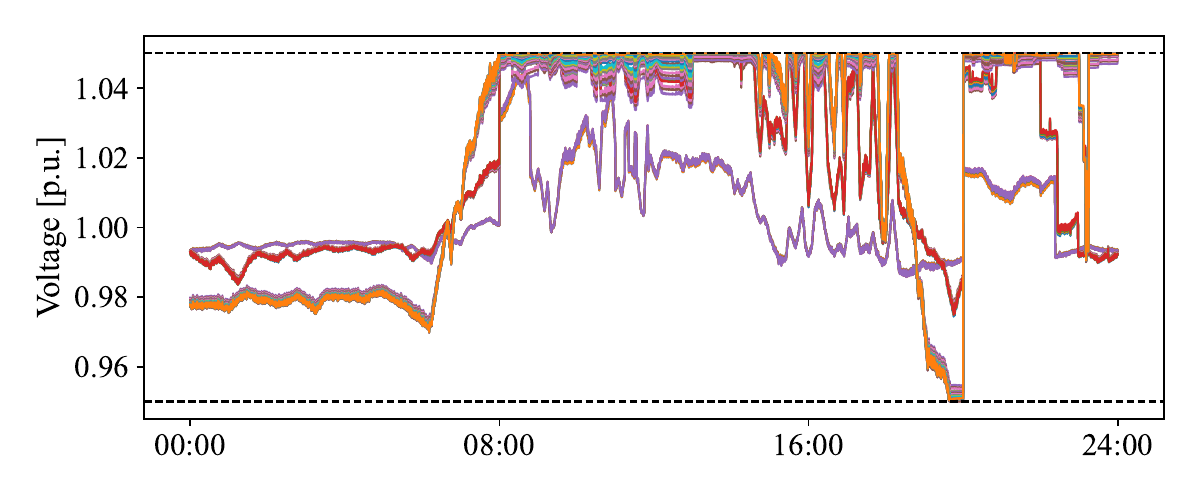}
    \caption{Voltage level for all nodes using the proposed controller}
    \label{fig:Voltage_results}
\end{figure}

Figure \ref{fig:E_exp} illustrates the import and export tariffs, along with the active power exported at the point of connection to the higher-level grid for the different control strategies. It can also be observed that the FA method exports significantly more when export tariffs are low, indicating higher PV production during these periods. Finally, Fig. \ref{fig:Voltage_results} demonstrates that the proposed strategy successfully maintains the voltage level at all nodes within the prescribed limits.

Even if accounting for future conditions brings the proposed aggregation method much closer to the optimal strategy, the remaining differences can be attributed to several factors. First, the OMNI strategy can anticipate future events and proactively discharge the BESSs at the best possible time. Second, the proposed aggregation method relies on polygonal approximations of the inverter's apparent power constraints. Third, the OP horizon was discretized into $N_C$ intervals. Fourth, the conservative formulation of the OP problem may limit the benefits of the BESS flexibility. Finally, the number of segments used to approximate the OP-stage output may also have a small effect.  

\subsection{Computation Time Analysis}
In the proposed configuration, the operational planning phase takes an average of 7.77 seconds to be solved every 15 minutes. This value may vary with the number of controllable assets and the number of intervals $N_C$. The projection and disaggregation steps are much faster, with average computation times of 0.42 seconds and 4 milliseconds, respectively. Being purely local, the duration of these steps is independent of the number of controllable nodes in the network. Still, it varies with the number of critical regions and, therefore, the number of flexible assets.

The most time-demanding step is the centralized problem resolution, which requires 0.80 seconds on average. This computation time is expected to increase slightly as the number of controllable nodes increases, and remains nearly independent of the number of local assets. 

These results demonstrate that the proposed method is compatible with real-time operating requirements.

\section{Conclusion}
\label{sect:conclusion}
This paper presents a novel approach to managing low-voltage distribution systems based on predictive flexibility aggregation through flexibility charts that capture feasible power injections and their associated values. Based on a set of parameters, the aggregated flexibility is precomputed using offline MPO. A parametric operational planning problem is proposed to account for the impact of current decisions on future costs. 

In real time, the result of the operational planning problem is integrated into the solution of the offline problem to tailor it to the current market period. Real-time measurements are then introduced to assess the flexibility charts.

By shifting most of the computational effort offline, the method significantly reduces the online computational burden and is compatible with real-time requirements. The approach was experimentally validated on a 43-bus system with realistic load and generation profiles. The results demonstrated that the proposed approach brought the FA approach much closer to the optimal solution, while relying solely on the current state of the system and the day-ahead local forecasts.

By aggregating the flexibility of individual assets, the proposed approach increases data privacy and simplifies the coordination of controls. It is also expected to scale efficiently and remain convenient for large systems, since most of the computations are performed locally or offline.

In future works, the proposed methodology will be extended to unbalanced three-phase systems and adapted to include other residential devices such as heat pumps and electric vehicles. Future work will also test this method in situations with incomplete measurement and imperfect communication.

\bibliographystyle{IEEEtran}
\bibliography{Bibliography} 
\end{document}